\documentclass[fleqn,usenatbib]{mnras}

\usepackage{newtxtext,newtxmath}
\usepackage[T1]{fontenc}
\usepackage{caption}
\DeclareRobustCommand{\VAN}[3]{#2}
\let\VANthebibliography\thebibliography
\def\thebibliography{\DeclareRobustCommand{\VAN}[3]{##3}\VANthebibliography}

\usepackage{graphicx}	
\usepackage{amsmath}
\usepackage{enumerate}
\usepackage[dvipsnames]{xcolor}

\title[Measuring mass functions in X-ray]{Measuring interacting binary mass functions with X-ray fluorescence}

\author[C. Dashwood Brown et al.]{
C. Dashwood Brown,
P. Gandhi,
P.A. Charles
\\
School of Physics and Astronomy, University of Southampton, Southampton SO17 1BJ, UK\\
}

\date{September 2022}

\pubyear{2022}

\begin{document}
\label{firstpage}
\pagerange{\pageref{firstpage}--\pageref{lastpage}}
\maketitle

\begin{abstract}
The masses of compact objects in X-ray binaries are best constrained through dynamical measurements, relying on radial velocity curves of the companion star. In anticipation of upcoming high X-ray spectral resolution telescopes, we explore their potential to constrain the mass function of the compact object. Fe K line fluorescence is a common feature in the spectra of luminous X-ray binaries, with a Doppler-broadened component from the inner accretion disc extensively studied. If a corresponding narrow line from the X-ray irradiated companion can be isolated, this provides an opportunity to further constrain the binary system properties. Here, we model binary geometry to determine the companion star’s solid angle, and deduce the iron line's equivalent width. We find that for systems with a mass ratio $q$\,>\,0.1, the expected K$\alpha$ equivalent width is 2–40\,eV. Simulations using XSPEC indicate that new microcalorimeters will have sufficient resolution to be able to produce K$\alpha$ emission line radial velocity measurements with precision of 5--40\,km\,s$^{-1}$, for source continuum fluxes exceeding $10^{-12}$\,erg\,cm$^{-2}$\,s$^{-1}$. Several caveats need to be considered; this method is dependent on successful isolation of the narrow line from the broad component, and the observation of clear changes in velocity independent of scatter arising from complex wind and disc behaviour. These issues remain to be proven with microcalorimeters, but this method has the potential to constrain binary parameters where optical measurements are not viable.
\end{abstract}

\begin{keywords}
X-rays: binaries -- techniques: spectroscopic -- black hole physics
\end{keywords}



\section{Introduction}
The mass of compact objects, i.e. neutron stars or black holes, is of great interest as this fundamental characteristic is linked to stellar evolution and populations, supernovae dynamics, and compact object formation. Additionally, understanding the event rates for gravitational waves relies upon understanding the mass distribution of black holes \citep{Abadie2010}. Despite extensive study, a number of questions related to stellar mass black holes remain, for example, the so called "mass-gap": the phenomena that most galactic compact objects are less than 3\(\textup{M}_\odot\) (neutron stars) or greater than 4.5\(\textup{M}_\odot\) (black holes). There is limited evidence of objects that have masses of between 3-5\(\textup{M}_\odot\) \citep{Bailyn1998,Ozel2010,Farr2010} - although the mass functions of some of these objects give ranges encompassing this gap \citep{Heida2017}. At present, there is no universally agreed upon explanation as to why this might be. This is often attributed to supernovae energetics \citep{Fryer2001}, while more recently some have argued that such small black holes may exist, but not exhibit typical features due to differences in the way they interact with their surroundings and are therefore only detected with LIGO/VIRGO observations \citep{Thompson2019,Cromartie2020,Jayasinghe2021}. Further study of the mass distribution of stellar mass black holes may be able to confirm or refute the existence of such a gap, helping to evolve black hole formation theories.

The masses of compact objects cannot be measured directly, however they can be constrained though observations of the radial velocity curves of the companion star \citep{frank2002, Casares2015}, both in non-interacting as well as interacting (X-ray) binaries. Here, we focus on interacting X-ray binaries (XRBs).

Consider an XRB system, comprising a compact object and a companion star, orbiting a centre of mass (defined by $M_1a_1 = M_2a_2$). Assuming circular orbits (a reasonable assumption, given that tidal effects typically circularize orbits on comparatively short timescales), the radial velocity amplitude of the companion star is given by:
\begin{equation}
    K_2 = \frac{2 \pi}{P} a_2 \mathrm{sin}i 
\end{equation}

Using the centre of mass definition and Kepler's third law, the `mass function' can be derived:
\begin{equation}
    f(M_1) = \frac{M_{1}^3\mathrm {sin} ^{3}i}{(M_{1}+M_{2})^{2}}={\frac {P_{\mathrm {orb} }\ K_2^{3}}{2\pi G}},
\end{equation}

where M$_1$ is the mass of the compact object, M$_2$ is the observed mass of the companion, $K_2$ is the peak Keplerian velocity of the companion and $P_{orb}$ is the orbital period. $f(M_1)$ provides an important  minimum mass for the compact object.

Mass functions have been derived for many black hole XRBs, with the radial velocity curve typically relying on optical or infrared spectra \citep{Casares1992,McClintock2001,Heida2017}. In the optical, Bowen fluorescence from the irradiated surface of the mass donor in Scorpius X-1 led to the first dynamical constraints on the system \citep{Casares2003}. It is important to note here the caveat of the `K-correction'; given that the emission comes from the irradiated ’half’ of the donor, this is offset from the donor’s centre of mass, this requires a small correction that needs to be accounted for when calculating mass functions \citep{MDarias2005}. Without the K-correction, this leads to a lower limit on the mass function. Thus far, there have been limited attempts to extend this method to other wavelength ranges. \citet{Zhang2012} used the Doppler shift in X-ray absorption features associated with an accretion disc to constrain the mass of the compact object and the inclination of GRO J1655--40.  In 2018, \citeauthor{Ponti2018} used absorption features in the X-ray spectra of a low mass X-ray binary (LMXB) to derive radial velocity measurements and therefore constrain the mass of the companion. Their findings were in agreement with previous estimates from other methods, indicating promise in the idea of expanding mass function methodologies to different wavebands. XRBs are a suitable testbed, and if successful, the same method could be applied to non-accreting active binaries. 

XRB X-ray spectra also often exhibit {\em emission} features. In particular, the fluorescent iron emission is prominent as a result of a high Fe abundance and fluorescence yield.  The K$\alpha$ emission feature at a rest energy of $\approx$\,6.4\,keV (or a wavelength of 1.9387 {\AA}) has been observed in a large number of XRBs, and is a potentially valuable means to derive system properties, especially the Doppler-broadened component that arises deep within the potential well of the binary close to the compact object  \citep{Basko1978,Makishima1986,Fabian1989,George1991, Ji2012}. 
Any narrow line component that might arise farther out from the core, with a FWHM of a few eV, is best suited to observations at high spectral resolution. 

 \citet{Torrejon2010} provide a comprehensive summary of this narrow component of K$\alpha$ detections in bright XRBs using {\em Chandra} gratings, finding them to be present in all high mass systems (HMXBs), and a small number of LMXBs. They attribute the source of the lines to be reprocessing of X-ray photons in cold stellar winds, arguing that this explains the lack of observations in low mass systems, which typically do not have substantial stellar winds. The equivalent widths (EWs) of these are large, sometimes exceeding 200\,eV, a fact that has interesting consequences for their origin, as we will discuss herein. Conversely, \citet{Torrejon2015} examine \emph{Chandra} observations of QV Nor, and present evidence that the main part of the Fe K$\alpha$ emission comes from the illuminated side of the donor, which is in agreement with our assumptions.

It is encouraging that narrow line components have already been successfully isolated and studied using gratings. But observations so far have, very likely, found only the `lowest hanging fruit,' with other, fainter and narrower components remaining to be detected. This is set to change with major upcoming advancements, including the highly anticipated launch of {\em XRISM} in 2023 \citep{xrism} \& {\em Athena} \citep{athena} in the next decade. With these, high-precision X-ray observations will be used to address a number of underlying questions related to accretion processes and compact objects \citep[e.g.,][]{XCALIBUR}. These will be enabled by novel microcalorimeter technologies, with an approximate order-of-magnitude improvement in spectral resolution around the important Fe K band energies $\sim$\,6--7\,keV, relative to best current grating capabilities \citep{Canizares2005}. It is therefore necessary to carry out modelling and simulations in preparation, in order to maximise their scientific output.

In this paper, we investigate the possibility of using observations of the K$\alpha$ emission line from the surface of a companion star to find the mass function of compact objects. We do this by considering the geometry of the binary system, and determining the apparent solid angle of the companion star. This allows us to estimate the intensity of the K$\alpha$ line relative to the background continuum and the equivalent width we expect to observe, and from this we carry out simulations to determine the accuracy to which we may be able to resolve the fluorescent line and measure radial velocity.

\section{Methods}
Our goal is to estimate the strength of the fluorescence component from the donor star in XRBs, and to explore the viability of utilising this to constrain mass functions. In order to determine the potential validity of this method, we must first determine the expected intensity profile of the K$\alpha$ emission from the secondary star within a binary system. This comprises three main terms; (i) the equivalent width expected from any part of the companion's surface, given the chemical composition of the reprocessing surface and the X-ray continuum flux; (ii) the solid angle presented by the companion star to the X-ray source; and (iii) the `projected area': a term encompassing the variations in equivalent width associated with the viewer (i.e. changes with phase, inclination, etc.). We therefore have:

\begin{equation} \label{eq:3}
    EW_{obs} = EW_I \times \Omega(q) \times \textit{$\alpha$}(q,i,\Phi)
\end{equation}

where $EW_I$ is the equivalent width of the K$\alpha$ line per unit area of reprocessing material (see below, and also \citealt{Sunyaev1998,Kallman2004}). $\Omega$ is the solid angle for a given system (a function of mass ratio, $q$), and $\alpha$ is the projected area to the observer (a function of $q$, inclination $i$, and orbital phase $\Phi$). Here, we note the work of \citep{Basko1978}, who also computed the expected equivalent width of the Fe K$\alpha$ line from donor stars. Our work is broadly consistent with this prior pioneering work, but improves upon it in various ways; e.g., \citeauthor{Basko1978} assumed a perfectly spherical companion star and also ignored the effects of disc shadowing that we discuss below. 

\subsection{Surface Composition}

The first of these terms is a function of the  composition of the reflecting surface, the continuum flux, and other characteristics intrinsic to the system and reprocessing material.

\begin{equation}
EW_I = N_{H} A_{Fe} Y \frac{\int_{E_{K}}^{\infty} E^{-2} \sigma_{ph} dE}{I(6.4)}
\end{equation}

where $N_{H}$ is the hydrogen column density (N$_H$ = 10$^{23}$--10$^{24}$\,atoms\,cm$^{-2}$); $A_{Fe}$ is the iron abundance relative to hydrogen (2\,$\times$\,10$^{-5}$); $Y$ is the fluorescence yield (typically 0.3); $I(E)$ is the photon flux spectrum of the X-ray source (assumed to be a power law, of form $E^{-2}$ photons\,s$^{-1}$\,cm$^{-2}$\,keV$^{-1}$); and $\sigma_{ph}$ is the photo-absorption cross-section for the iron K-shell for photons of a given energy ($\sigma_{ph}(E) = 3.5 \times 10^{-20} \times \frac{E_{K}^{3}}{E^{3}}$ cm$^2$ atom$^{-1}$) \citep{Sunyaev1998}. E$_K$ is the iron K-shell absorption energy ($\approx$\,7.11\,keV), and $EW_{I}$ is typically of order 30--300\,eV \citep{Basko1978, Sunyaev1998, Torrejon2010}.

\subsection{Solid Angle}
For closely separated binaries, the orbiting stars are distorted such that they can no longer be approximated as spherical. The outer layers remain gravitationally bound to their star, but experience the effects of the compact object, leading to stars with a `tear-drop' shape \citep{kopal1959}. This can be described using Roche lobe geometry. We can solve the Roche potential equation to determine $R$ (radius of the star) for any given $\theta$ and $\phi$ (angular position from centre of companion). This gives us a complete picture of the geometry of the distorted star, and from this we can determine the solid angle. 

The solid angle defines the field of view that a given object (in this case, the companion star) covers as viewed from the X-ray source \citep{Eriksoon1990}. Measured in steradians, the solid angle is given by: $\Omega ={\frac {A}{r^{2}}}$ (where $A$ is the apparent surface area of the companion star and $r$ is the distance from the compact object) and it is therefore necessary to find the two-dimensional surface area of the companion star as seen by the compact object.

Consider the derivation of the surface area of a sphere; we define an infinite number of segments, each with dimensions $rd\theta$ and $rsin\theta$, and integrate the areas of these segments over the whole surface.

By substituting constant \emph{r} with a function of $\theta$ and $\phi$, the reflecting surface of a distorted companion star can be found in the same way. Given symmetry about the \emph{x}-axis, the radius at any given $\phi$ will be the same for all values of $\theta$, and so by taking $\theta = \pi/2$ we find:
\begin{equation}
     \Phi_{S} = \frac{1}{r}+q\left(\frac{1}{\sqrt{1-2r\cos\phi +r^2}} - r\cos\phi\right)+\frac{q+1}{2}r^2  
\end{equation}
where $\Phi_{S}$ is the potential at the stellar surface and $q$ = M$_2$/M$_1$. The above can be solved to find the radius at a given $\phi$, and we therefore have: 
\begin{equation}
    \Omega = \int_{}^{} R(\phi)^2 \sin\phi d\phi 
\end{equation}
where \emph{R} is a function of $\phi$ and $q$ only.

\begin{figure}
\begin{minipage}{.3\columnwidth}
  \centering
  \includegraphics[width=2.8cm]{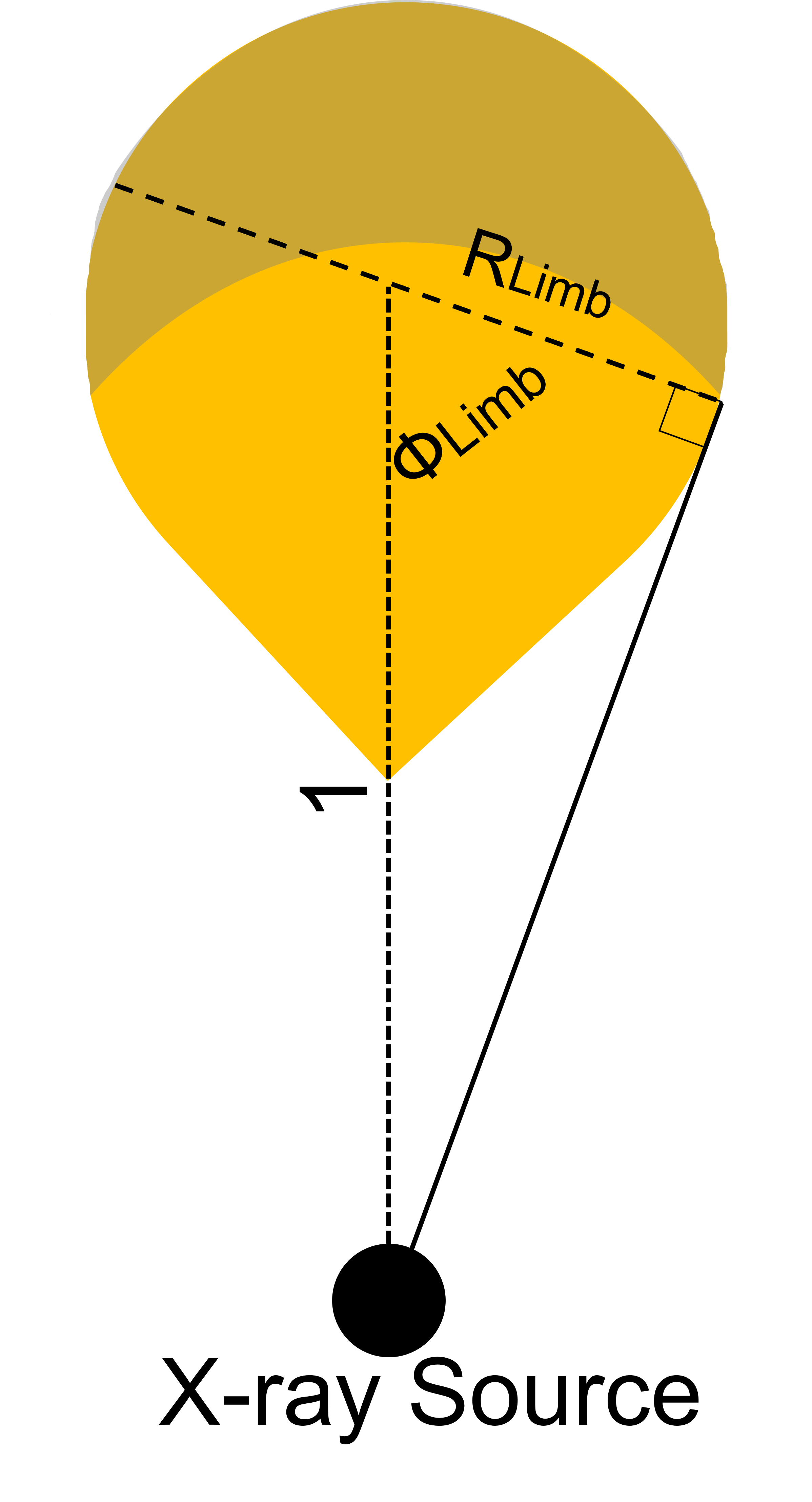}
  \label{fig:limb1}
\end{minipage}%
\begin{minipage}{.7\columnwidth}
  \centering
  \includegraphics[width=6cm]{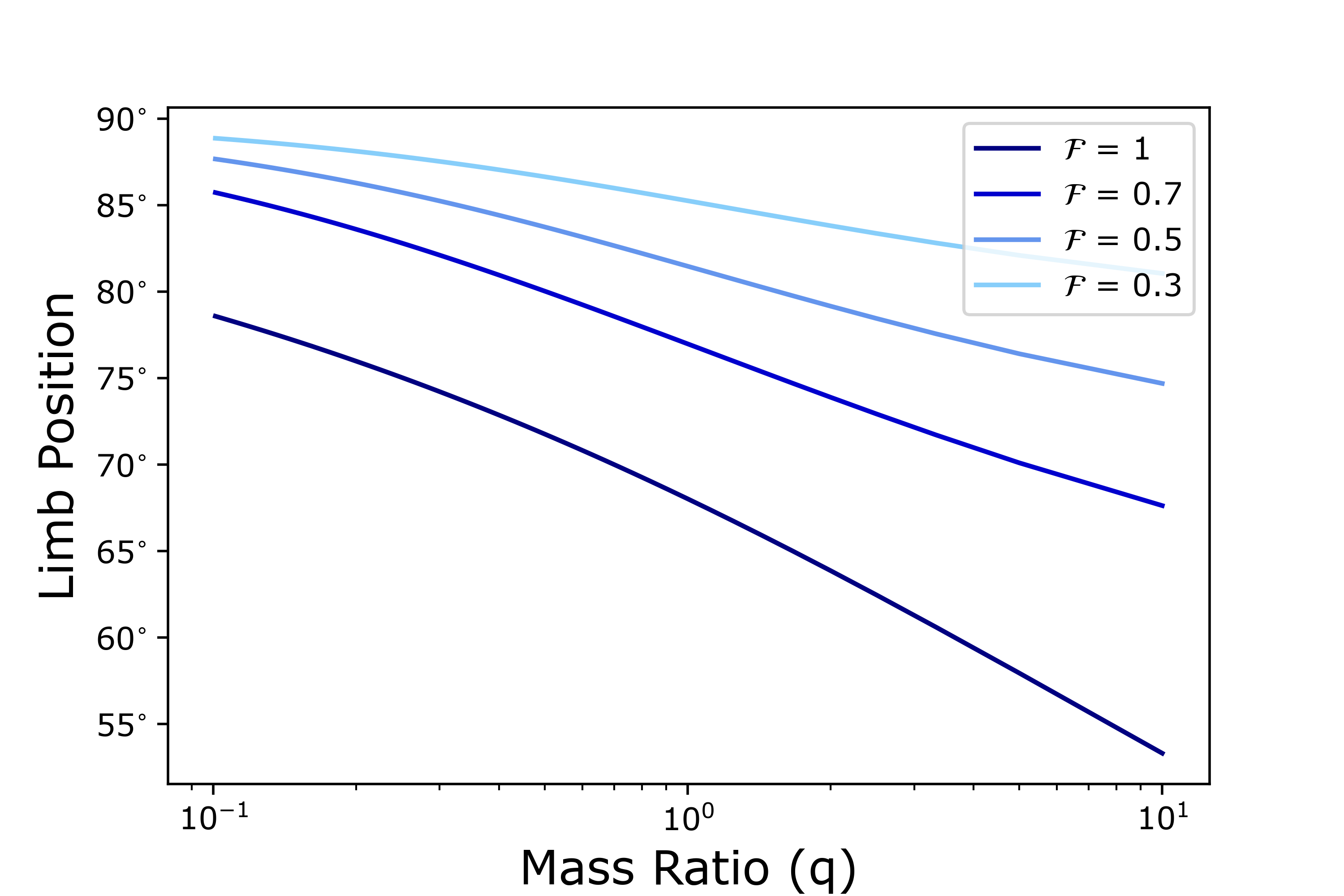}
  \label{fig:limb2}
\end{minipage}
 \caption{({\em Left}) Illustration of limb position as seen from compact object; ({\em Right}) limb position (expressed as angle $\phi$ from $x$-axis) as a function of mass ratio for various filling factors $\mathcal{F}$.}
 \label{fig:Limb}
\end{figure}

To find the reflecting surface area at a given phase, the above integral can be carried out, within the limits that define the reflecting region. We find the limits by considering the geometry of a point source emitting light onto the star, illuminating the area within the limbs, as shown in Fig.\,\ref{fig:Limb}. The limbs of the star are defined by the point at which a line subtended from the surface of the star to the point source (i.e. a compact object) is perpendicular to a line from the centre of the star to the surface (\citealt{Hamme1993} and references therein); i.e. where $\cos\phi_L$ = R($\phi_L$) (given that our radius values are expressed in terms of $a$).

Therefore, our integral is from $-\phi_{L}$ to $\phi_{L}$ (where $\phi_{L}$ is typically 60-80$^{\circ}$ - Fig.\,\ref{fig:Limb}(right) shows a plot of limb position (degrees) as a function of mass ratio). This gives us the surface area of the star that is illuminated by a source directly in front of it (at phase 0), and is therefore the maximum reflecting area, and the solid angle presented to the source X-ray flux by the companion star. 

\subsection{Disc Shadowing}
\begin{figure}
\begin{minipage}{\columnwidth}
\includegraphics[width=\linewidth]{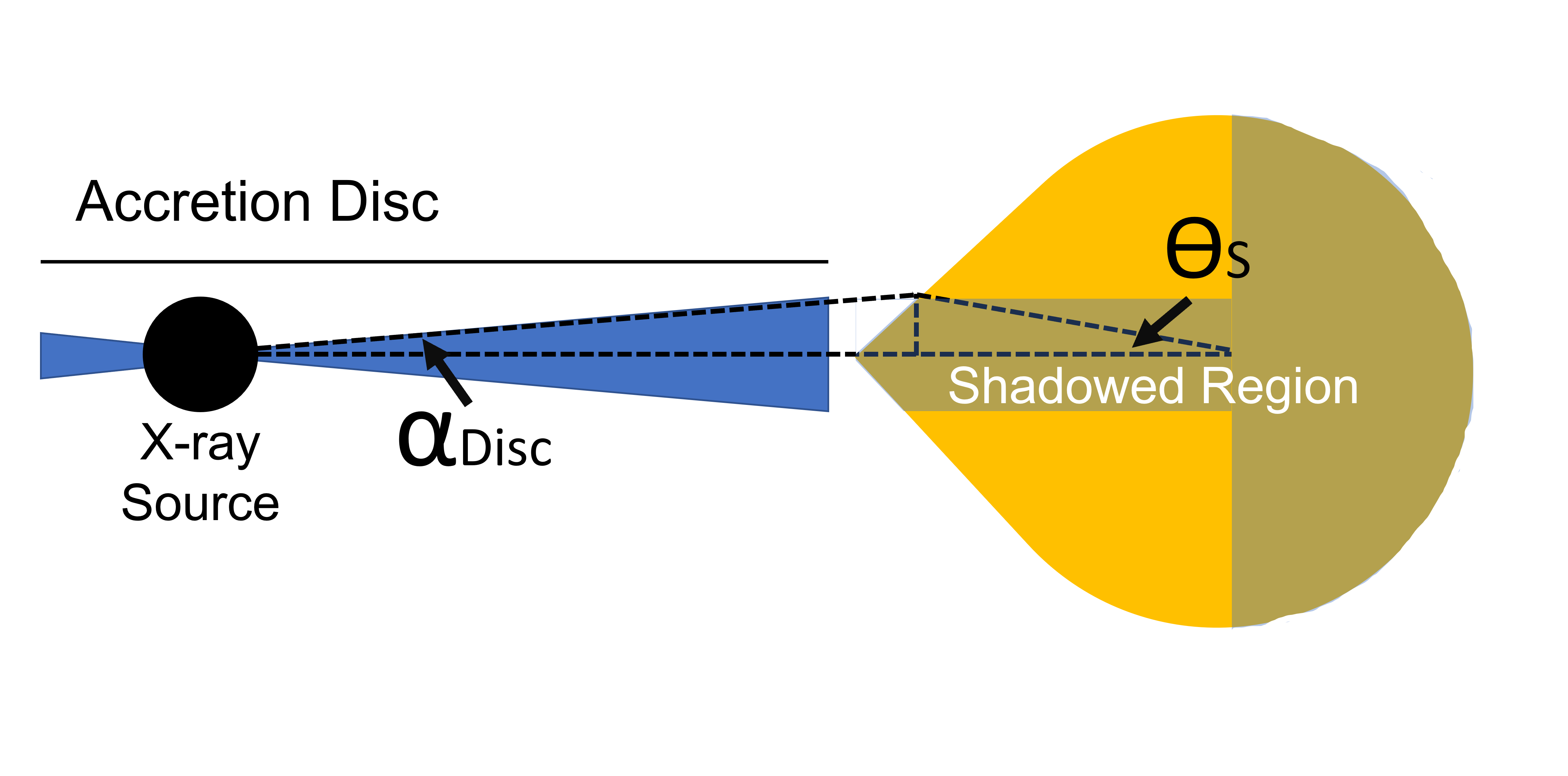}
\end{minipage}
\begin{minipage}{\columnwidth}
\includegraphics[width=\linewidth]{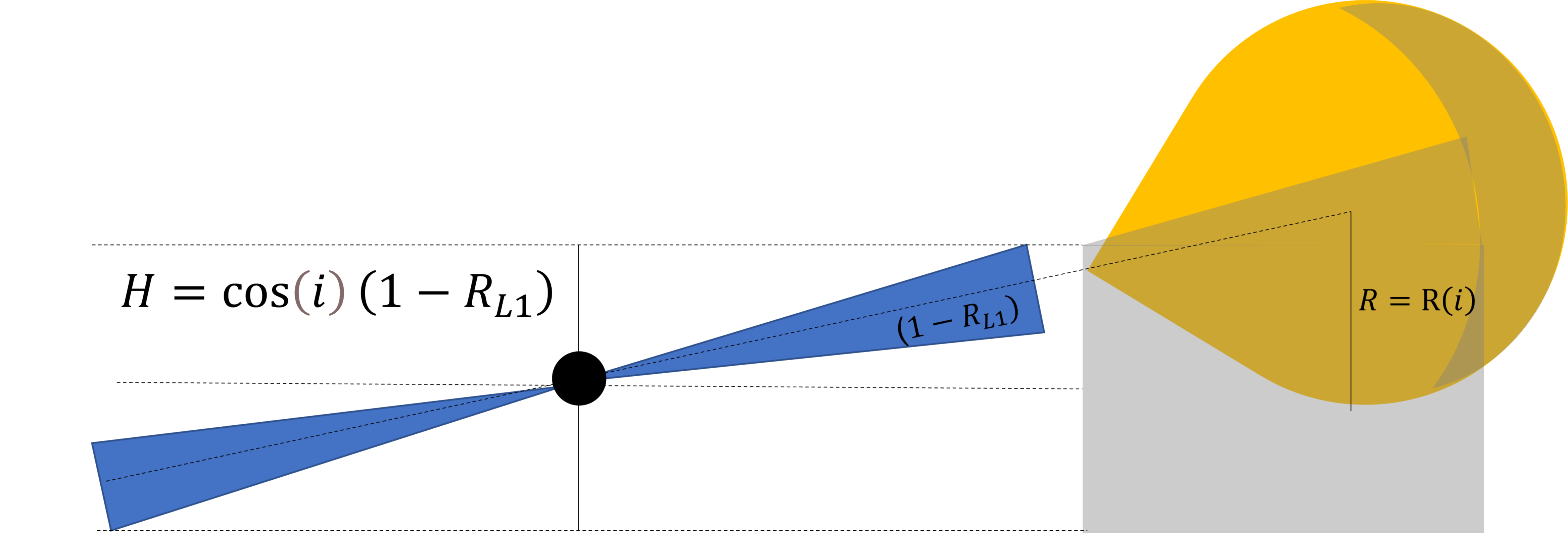}
\end{minipage}
\caption{Schematics of the shadow cast by an accretion disc in a LMXB, as seen for an inclination of ({\em upper}) 90$^\circ$ and ({\em lower}) $\sim$75$^\circ$. The observer is located towards the left (i.e. the donor is at superior conjunction). For an HMXB, the companion star is sufficiently large that the shadow cast by the accretion disc is comparatively much smaller.}
\label{fig:shadow}
\end{figure}

Where the above gives a description of the solid angle of the companion star in a binary, we must also consider the effects of an accretion disc that surrounds the compact object. An accretion disc will have the effect of casting a shadow over the companion star, such that there is a region on the surface that will not exhibit X-ray reflection from the point of view of an observer. Fig.\,\ref{fig:shadow} shows this shadowing, for (a) a system at 90$^\circ$ inclination, and (b) a system at some other non-zero inclination. The shaded regions on the companion star are those which will not exhibit reflection, either because they lie beyond the limbs from the point of view of the emitting compact object, or because they are enshadowed by the accretion disc. This effect is minimal for HMXBs, where the companion star is significantly larger than the accretion disc. Conversely, the donors in systems where $q$ < 0.01 are so small such that they are totally in the shadow of the disc, and see no X-radiation at all. This is dependent on outer disc elevation or flaring angle, e.g. \citet{Jong1996} derives an average flaring angle of 12$^{\circ}$ for LMXB.
This effect can be approximated with some assumptions about the nature of the disc. If we use the equation from \citet{frank2002}:
\begin{equation}
    D_H \cong C_{s}(GM)^{-0.5}R_D^{1.5}
\end{equation}
where $D_H$ is the disc height, C$_s$ is sound speed ($\sim$1000\,km\,s$^{-1}$ ), and M is set as 5\(M_\odot\). By substituting $R_D$ (radius of the disc) as 1-$r_{L1}$ where $r_{L1}$ is the radius at the most elongated point of the star (the L$_1$ point), this gives us the thickness of the disc at the outer edge, toward the companion star (i.e. angle $\alpha_D$ in Fig.\,\ref{fig:shadow}).

By considering the geometry of the system, it is possible to find the area of the shadowed region;
$\theta_{S}$ is the angle that defines where the shadow begins on the stellar surface, found from solving for a given disc angular thickness: 
\begin{equation}
 \mathrm{tan}\alpha_{D} = \frac{x}{y} = \frac{R(\theta_{S})\mathrm{cos}\theta_{S}}{1-R(\theta_{S})\mathrm{cos}\theta_{S}}    
\end{equation}
The area of the shadowed region is given by:
\begin{equation}
    A = 2R(\theta_S)r_{Limb}\mathrm{sin}\theta_S
\end{equation}
and can be subtracted off the solid angle to find the re-processing area.

All of the above can be adapted to consider different inclinations. The nature of the distortion means that the solid angle of the companion star will change depending on the inclination of the system (i.e. the profile of the reflecting area will not change with phase when the inclination of the system is 0). Furthermore, at higher inclinations, additional phases of the star will be obscured by the disc: for example, at phase 0 or 1 the disc will have a much larger apparent "area" from the point of view of an observer, and shield more of the companion star, reducing the reprocessing area (as shown in Fig.\,\ref{fig:shadow}b, where the apparent height of the disc from the point of view of the observer results in a much larger portion of the companion's reflecting surface being obscured). We can incorporate this effect, by calculating the size of the disc from the point of the observer, and comparing this to the reflecting area. 

The combined effects of the accretion disc and the distortion of the companion star due to the Roche potential means that the solid angle presented by the companion to an observer will change with phase, resulting in a light curve with ellipsoidal variations.This is a purely geometrical effect, however there may be additional drivers of further variation in equivalent width, such as changing ionisation levels or $N_H$, as we discuss in Section 5.

\subsection{Projected Area $\alpha$}
The intensity of reprocessed flux from a companion star is a function of both solid angle and albedo, the latter of which comprises the effects of molecular composition and the angular distribution of radiation that is emitted from the surface \citep{Wilson1990,Coakley2003}. The effects of molecular composition are included in EW$_I$, however this additional term describes the geometric albedo, i.e. the effect that observation geometry has on the projected area of the companion star, which will depend on $q$, inclination, and phase. It is therefore necessary to consider the angles between the K$\alpha$ emitted flux and an observer.

We assume that emission from the surface of the star is isotropic at each small surface element, and therefore $\theta$ is the angle normal to the surface element. This can be calculated by considering that the Roche potential equations describe a plane surface, and the normal of any surface is defined by $f(x,y,z)=\nabla f$. We can therefore find the angles $\theta, \phi$ that define the unit normal vector coming off each point of the stellar surface from the gradient of the Roche potential equation.

$\frac{\partial f}{\partial \theta}\times \frac{1}{|\nabla f|}$ gives the angle $\theta_N$ that defines the normal to any point on the stellar surface, and by integrating over the illuminated surface, we find the proportion of K$\alpha$ emission that reaches the observer (i.e. taking the cosine to find the component of light that is directed towards an observer):

\begin{equation}
    \alpha = \int_0^{2\pi} d\Phi \iint_{S_{R}} \mathrm{cos}\left(\theta-\theta_N\right)\mathrm{sin}\theta\, \mathrm{cos}(\phi-\phi_N)\,d\phi\,d\theta
\end{equation}

We include the term:
\begin{equation}
    \mathrm{cos}\Theta = -\mathrm{sin}i\,\mathrm{cos}(2\pi\Phi)
\end{equation}
from \citet{Basko1978} to account for varying inclinations and phases. Here, $\Theta$ is the angle between two rays emerging from the
center of the companion, one pointing to the X-ray source, and the other to the observer. $\Phi$ is orbital phase.

We therefore have $\alpha$ as a function of phase and inclination, indicating the proportion of emission that is directed towards an observer.

\subsection{Expected Equivalent Width}
All of the above allows us to determine the expected equivalent width of the K$\alpha$ line produced from the companion star with respect to the continuum reflected off the disc (Equation \ref{eq:3}). The final outcome of the equivalent width for each orbital phase is based on the mass ratio $q$, filling factor $\mathcal{F}$, and inclination $i$ only - as the solid angle derived from the potential equations are in terms of orbital separation, by definition, and thus this system parameter cancels out. 

We can then plot the expected iron line equivalent width as a function of orbital phase, for systems of a given mass ratio \& inclination. 

\section{Results}
\begin{figure}
\begin{minipage}{\columnwidth}
 \includegraphics[width=9cm]{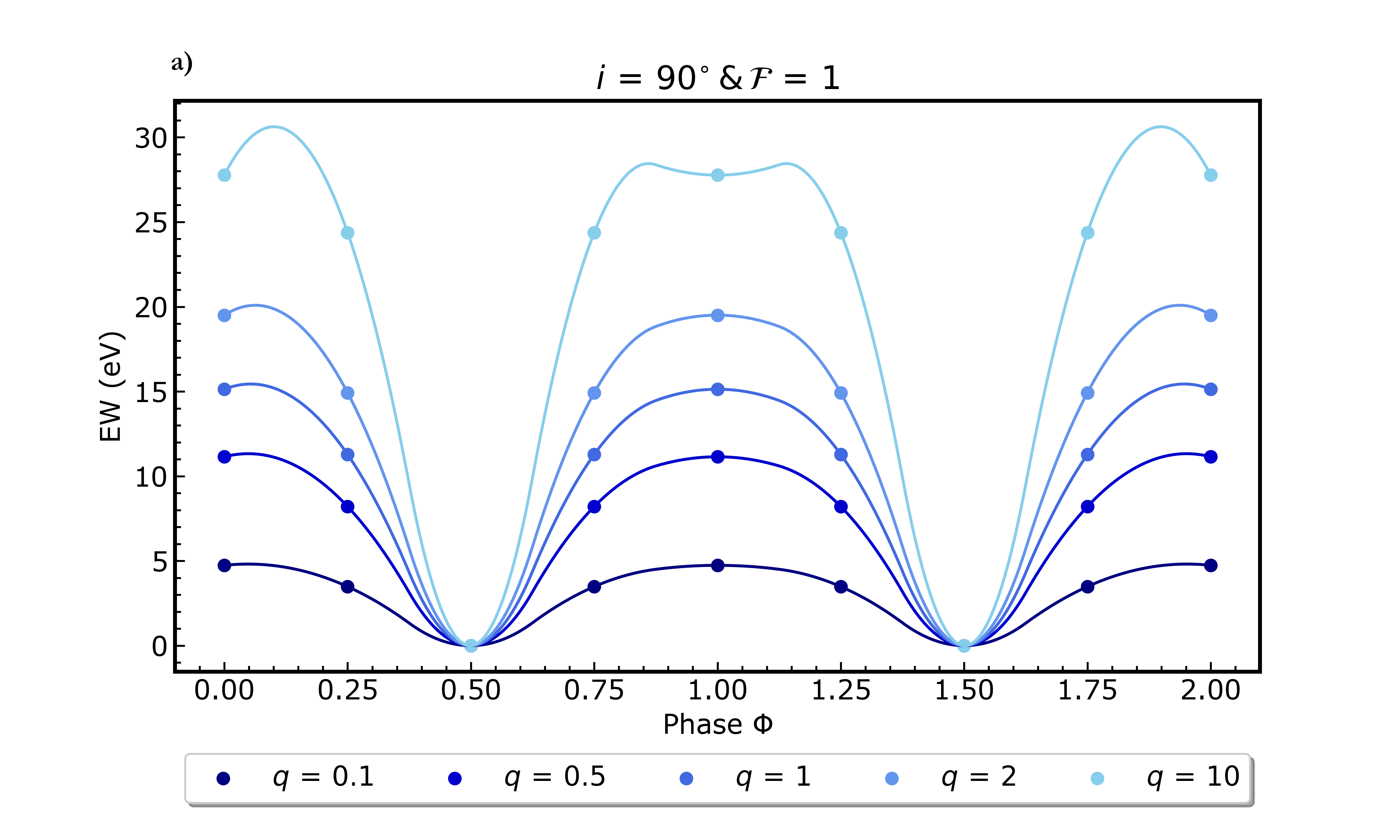}
\end{minipage}
\begin{minipage}{\columnwidth}
 \includegraphics[width=9cm]{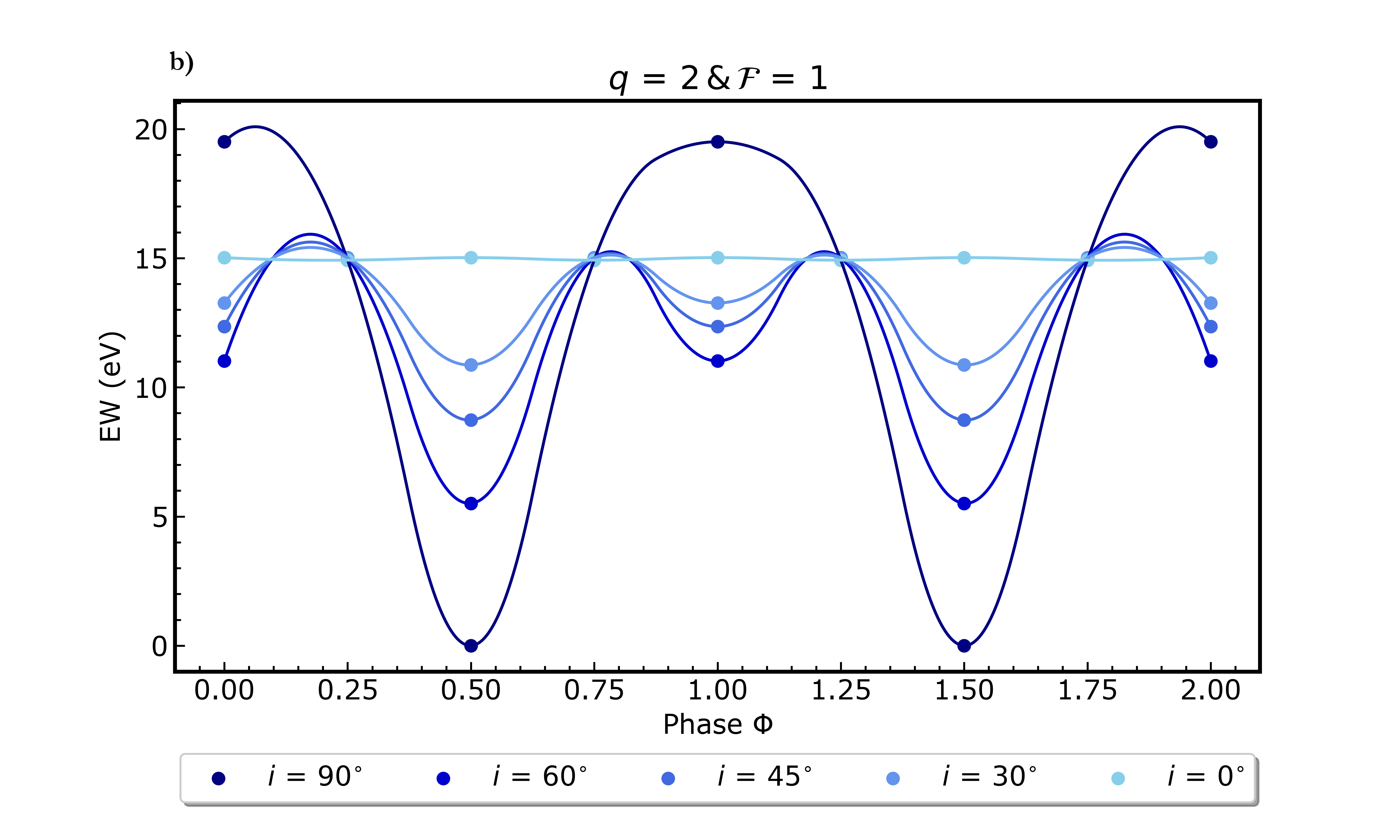}
\end{minipage}
\begin{minipage}{\columnwidth}
 \includegraphics[width=9cm]{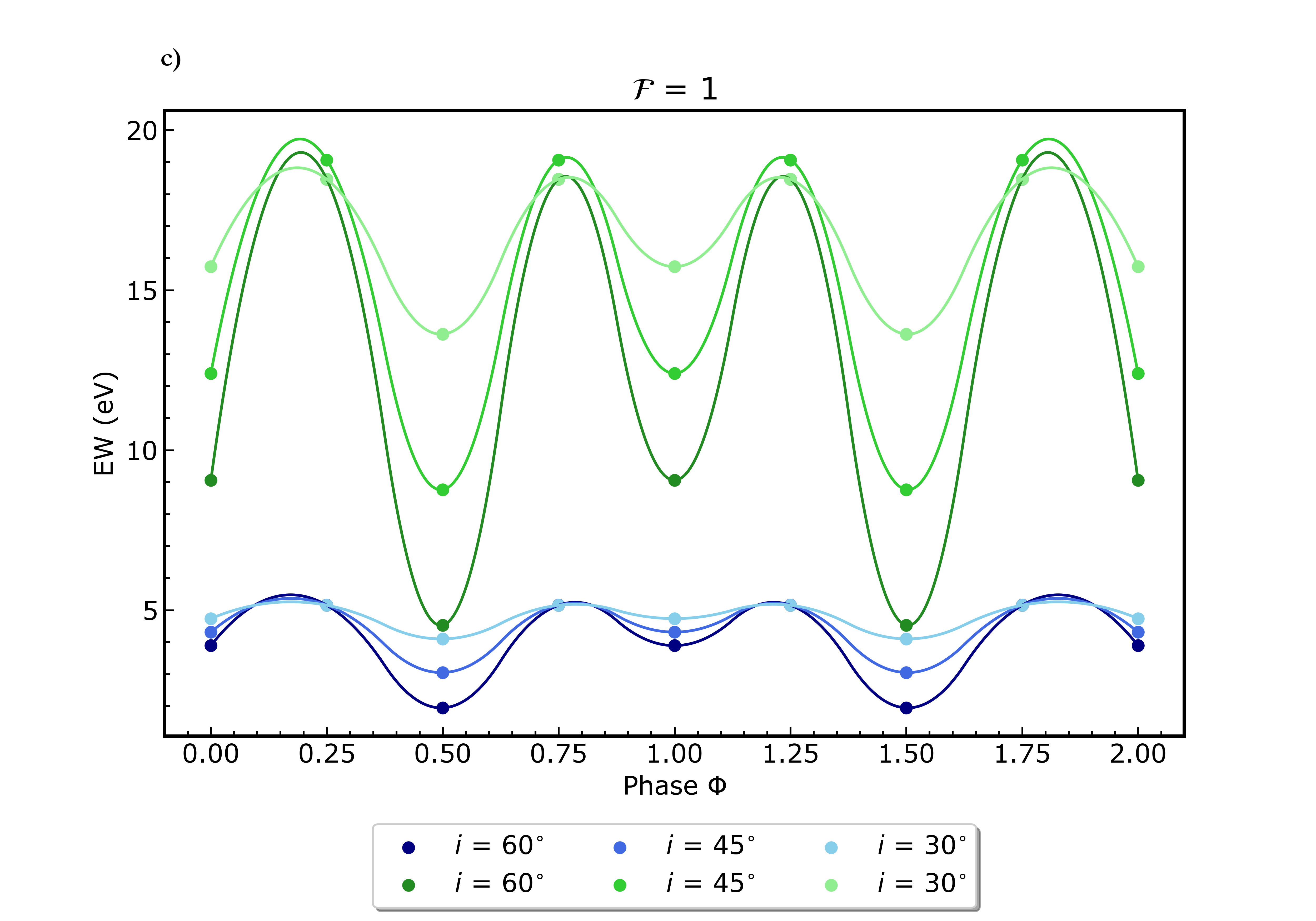}
\end{minipage}
\caption{The K$\alpha$ line equivalent width as a function of phase for ({\em a}) various mass ratios, ({\em b}) various inclinations, and ({\em c}) for systems with common inclinations and mass ratios of 0.2 (blue) \& 4 (green).}
\label{fig:ewcurve}
\end{figure}

We present plots showing the expected equivalent width of the K$\alpha$ line as seen by the observer (Fig.\,\ref{fig:ewcurve}). The presence of an accretion disc and the Roche distortion of the companion star results in significant ellipsoidal variations of intensity as a function of orbital phase.  Such light curves of XRBs (specifically Cyg\,X-3, Her\,X-1 and 4U\,1700-37) were first calculated by \citet{Basko1974}.

This ranges from approximately 2--25\% of EW$_{I}$, depending on mass-ratio and phase (Fig.\,\ref{fig:ewcurve}a); unsurprisingly, systems with a low $q$ (i.e. $q$\,$<$\,1, indicating a smaller companion star) reflect only $\sim$10$\%$ or less of the source emission, and the K$\alpha$ EW is less than 15\,eV. For systems with $q$\,$>$ 1, the observed EW is of order 15-30\,eV (10-30\% of EW$_{I}$), due to the significant stellar surface area and solid angle. A caveat here is that most plots are presented assuming a totally filled Roche lobe (i.e. $\mathcal{F}$\,=\,1) - it is currently unclear exactly what conditions lead to the Roche lobe being filled, however it can be argued that a larger $q$, and therefore a smaller compact object relative to the companion may mean it is less likely to be totally filling it's Roche lobe. For filling factors $\mathcal{F}\,<\,$ 1, the solid angle of the system will be smaller than for a totally filled Roche lobe, which would decrease the observed EW.

If we consider HMXBs (i.e. systems with $q$\,>\,2) we expect to see EW up to 20-30\,eV. The distribution of XRB inclinations indicates a majority have $i \sim$ 30--60$^{\circ}$, meaning these values decrease (to around 50--80$\%$ of their maximum value) but remain of the same order of magnitude (Fig.\,\ref{fig:ewcurve}b).

Fig.\,\ref{fig:ewcurve}c shows a plot of the observed EW for two distinct mass ratios ($q$\,=\,0.2 \& $q$\,=\,4, representing a typical LMXB and HMXB, respectively), and at 3 different inclination angles (i.e. 60$^\circ$, 45$^\circ$, and 30$^\circ$).  These clearly show the ellipsoidal variations with phase due to the distortion of the companion. Inclination has a greater effect on systems with larger $q$, but the observed EW is expected to be of the same order of magnitude, regardless of inclination.

XRB source emission can vary significantly in intensity. During quiescence, fluxes are typically of the order $10^{-10}$ -- $10^{-12}$\,erg\,cm$^{-2}$\,s$^{-1}$, but during outburst they can increase by 2 to 3 orders of magnitude, or even brighter in some instances.  The narrow-line K$\alpha$ emission from the companion star is expected to have an EW with respect to the background continuum reflected off the disc of 2-40\,eV. Having deduced the expected EW of narrow line K$\alpha$ emission, we go on to consider whether this line could then be resolved and tracked so as to constrain the mass function.

\section{Simulations}
Simulations were carried out using XSPEC to determine if {\em XRISM} will be able to resolve the K$\alpha$ fluorescence line, and estimate the errors in energy and therefore velocity. This allows us to determine the potential validity of using the Doppler shift in this line to find the mass function, and thereby constrain the mass of the compact object.

We create a number of models, each comprising a power law for background continuum plus a Gaussian, centred on 6.4\,keV, with a width 5\,eV. The continuum flux of these models ranges from F$_{2-10\,keV}$ = $10^{-8} - 10^{-12}$\,erg\,cm$^{-2}$\,s$^{-1}$, in order of magnitude increments. For each of these fluxes, two equivalent widths are modelled: 20\,eV and 4\,eV. 
Using the {\em XRISM} high resolution (7\,eV) response files \citep{xrismqrg} and FAKEIT, we create simulated spectra to determine the fluxes under which we may be able to detect and resolve the iron-line, and its associated errors.

\begin{figure}
\begin{minipage}{\columnwidth}
 \includegraphics[width=9cm]{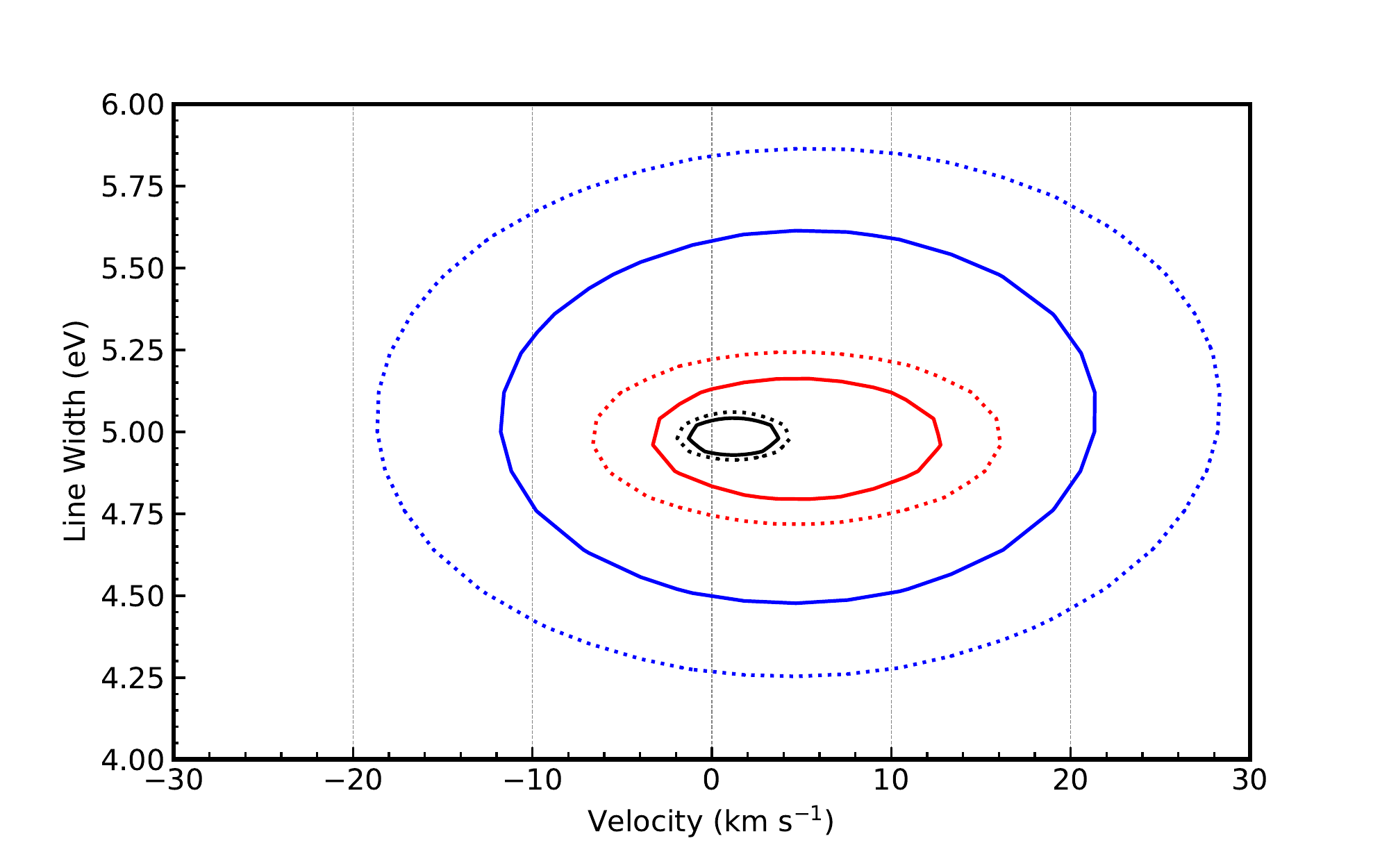}
\end{minipage}
\begin{minipage}{\columnwidth}
 \includegraphics[width=9cm]{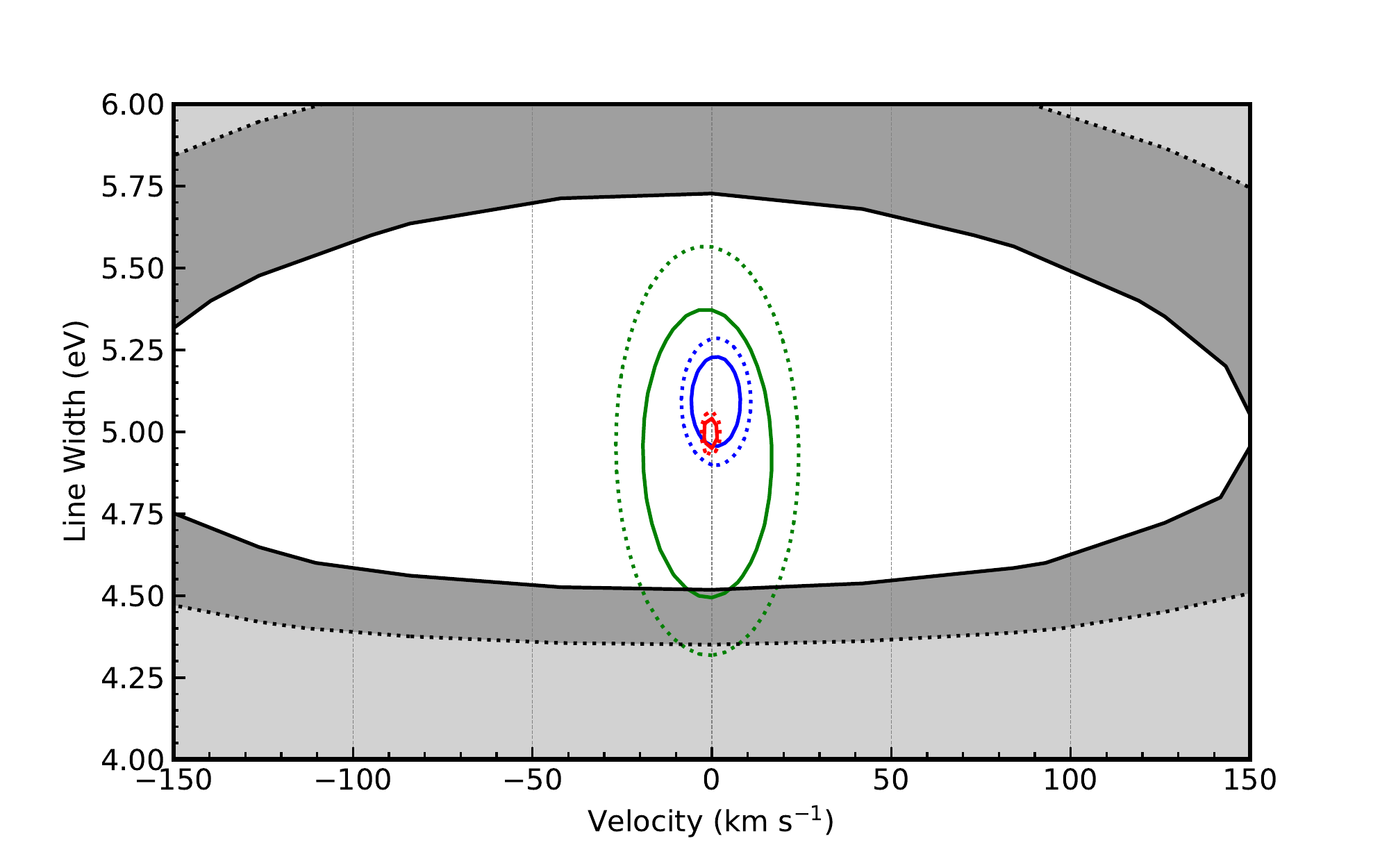}
\end{minipage}
\caption{Contour plot of $\chi^2$ confidence levels for the two fitting parameters (line energy and width), with line energies converted to velocities. 1-$\sigma$ and 2$\sigma$ confidence intervals are plotted in solid and dashed lines respectively, for fluxes of  $F$\,=\,$10^{-8}$\, (black); $F$\,=\,$10^{-9}$\, (red); $F$\,=\,$10^{-10}$\, (blue); and $F$\,=\,$10^{-11}$\,erg\,cm$^{-2}$\,s$^{-1}$ (green), and for EW\,=\,4\,eV ({\em upper}) and EW\,=\,20\,eV ({\em lower}). The shaded regions in the lower panel indicate velocity contours from a fit using {\em XMM-Newton} at $F$\,=\,$1\times10^{-9}$\,erg\,cm$^{-2}$\,s$^{-1}$.}
\label{fig:contour}
\end{figure}

Fig.\,\ref{fig:contour} shows a contour plot with 1-$\sigma$ and 2-$\sigma$ confidence intervals in velocity measurements (converted from line energy) against line-width, for three different flux levels ($10^{-8}$, $10^{-9}$, and $10^{-11}$\,erg\,cm$^{-2}$\,s$^{-1}$).

These simulations indicate that for higher fluxes (i.e. during outbursts, F$_{2-10\,keV}\sim\,$10$^{-8}$ -- 10$^{-10}$\,erg\,cm$^{-2}$\,s$^{-1}$), if {\em XRISM} is able to detect and resolve the narrow line, the errors in energy will be $0.01\,-\,0.1\,$eV. This corresponds to velocity errors for the companion star of 5--40\,km\,s$^{-1}$. In the best case scenario (outbursts of $10^{-8}-10^{-9}$\,erg\,cm$^{-2}$\,s$^{-1}$ and $q$\,$>$\,4) we may be able to resolve the K$\alpha$ line with velocity errors within a few \,km\,s$^{-1}$ . This relies on exposure times of 100\,ks; but further simulations were carried out with exposures of 50\,ks and 20\,ks, which show that, if the X-ray flux is bright (i.e. F$_{2-10\,keV}$ = $10^{-8} $--$ 10^{-9}$\,erg\,cm$^{-2}$\,s$^{-1}$) velocity errors remain well-constrained. 

The 1-$\sigma$, 2-$\sigma$ uncertainties as shown in Fig.\,\ref{fig:contour} are given in the Table 1. Note, higher fluxes rely on exposure times exceeding 50\,ks - this is therefore considered the limiting flux for which we may calculate radial velocity measurements: 

\begin{center}
\begin{tabular}{||c c c||} 
 \hline
 F$_{2-10\,keV}$ & 1-$\sigma$ & 2-$\sigma$ \\ [0.5ex] 
erg\,cm$^{-2}$\,s$^{-1}$ & km\,s$^{-1}$ & km\,s$^{-1}$\\
 \hline\hline
 $10^{-8}$ & 2 & 3  \\ 
 \hline
 $10^{-9}$ & 5 & 9 \\
 \hline
 $10^{-11}$ & 19 & 24 \\
 \hline
 $10^{-12}$ & 40 & 46  \\ [1ex] 
 \hline
\end{tabular}
\captionof{table}{1-$\sigma$ and 2-$\sigma$ errors in velocity for various X-ray fluxes, given 50\,ks exposure time.}
\end{center}

Typical HMXB companion star velocities range from 10 --100 \,km\,s$^{-1}$, meaning that our velocity errors may be of order 5-10$\%$.  However, given sufficient orbital phase coverage, it should still be sufficient to yield a radial velocity curve, from which the amplitude and period can be determined. 

\begin{figure}
    \centering
    \includegraphics[width=9cm]{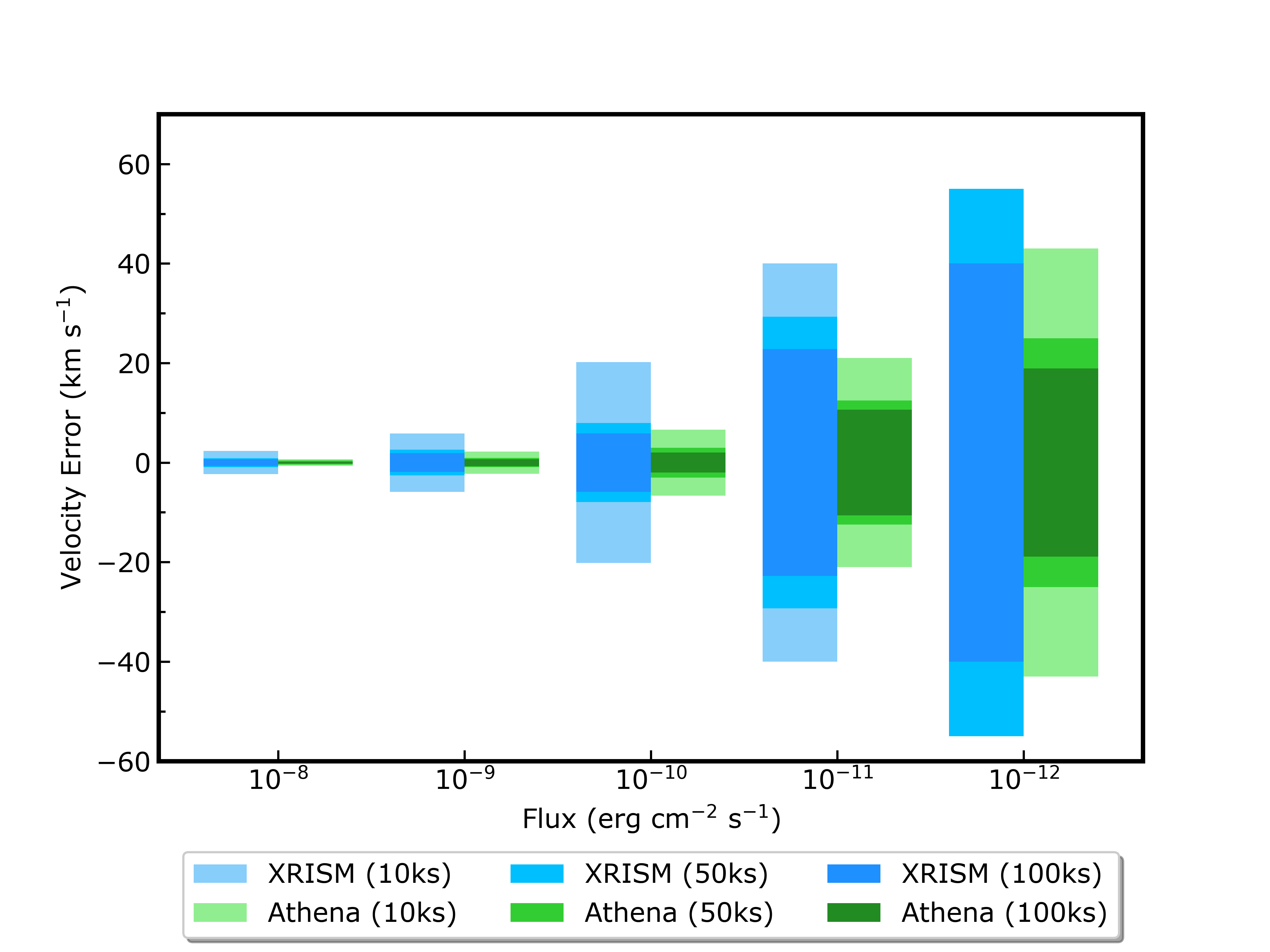}
    \caption{Statistical ($1-\sigma$) errors in epoch velocity determination with both XRISM and Athena as a function of source flux, for exposure times of 10\,ks, 50\,ks, and 100\,ks and an emission line of EW 4\,eV.}
    \label{fig:fluxerr}
\end{figure}
Fig.\,\ref{fig:fluxerr} shows the expected errors in velocity for a range of fluxes, using both the {\em XRISM} and {\em Athena} response files, and for different exposure times. 

\subsection{Simulated Velocity Curves}
Having calculated the equivalent width as a function of phase, we carried out simulations in XSPEC mimicking the spectra we expect to observe from the companion star, to determine how well line energies and radial velocities will be constrained.

\begin{figure}
\begin{minipage}{\columnwidth}
 \includegraphics[width=9cm]{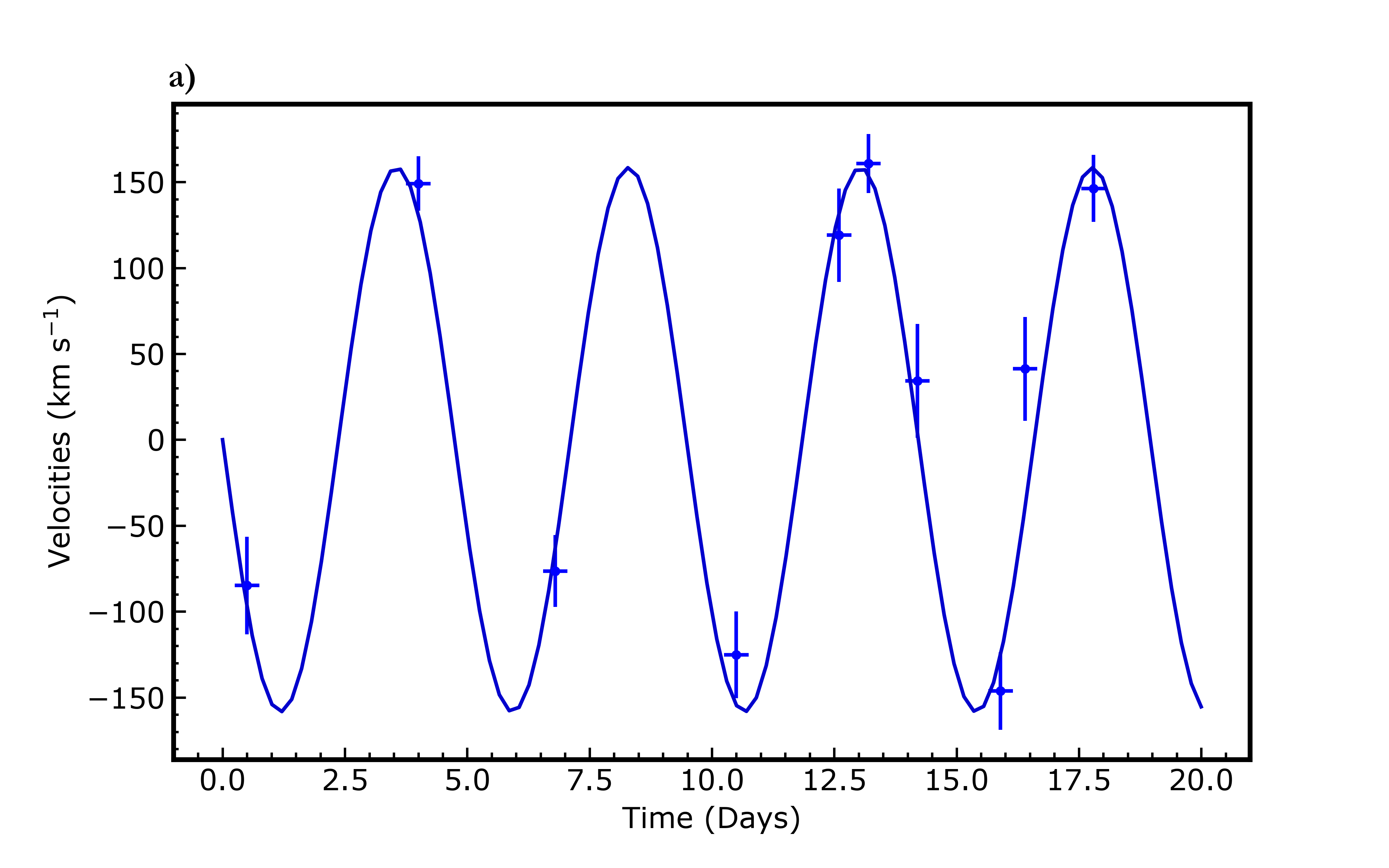}
\end{minipage}
\begin{minipage}{\columnwidth}
 \includegraphics[width=9cm]{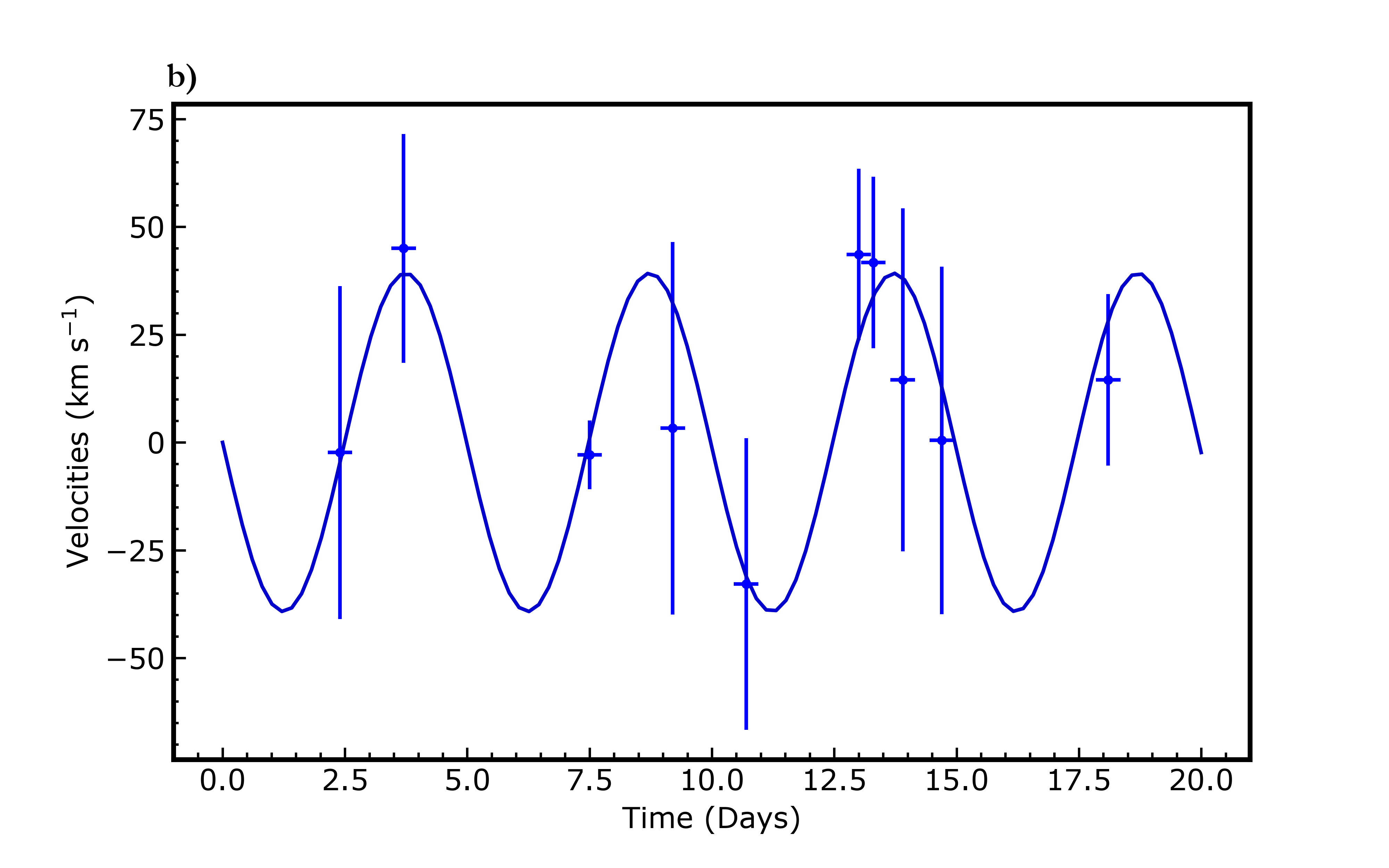}
\end{minipage}
\begin{minipage}{\columnwidth}
 \includegraphics[width=9cm]{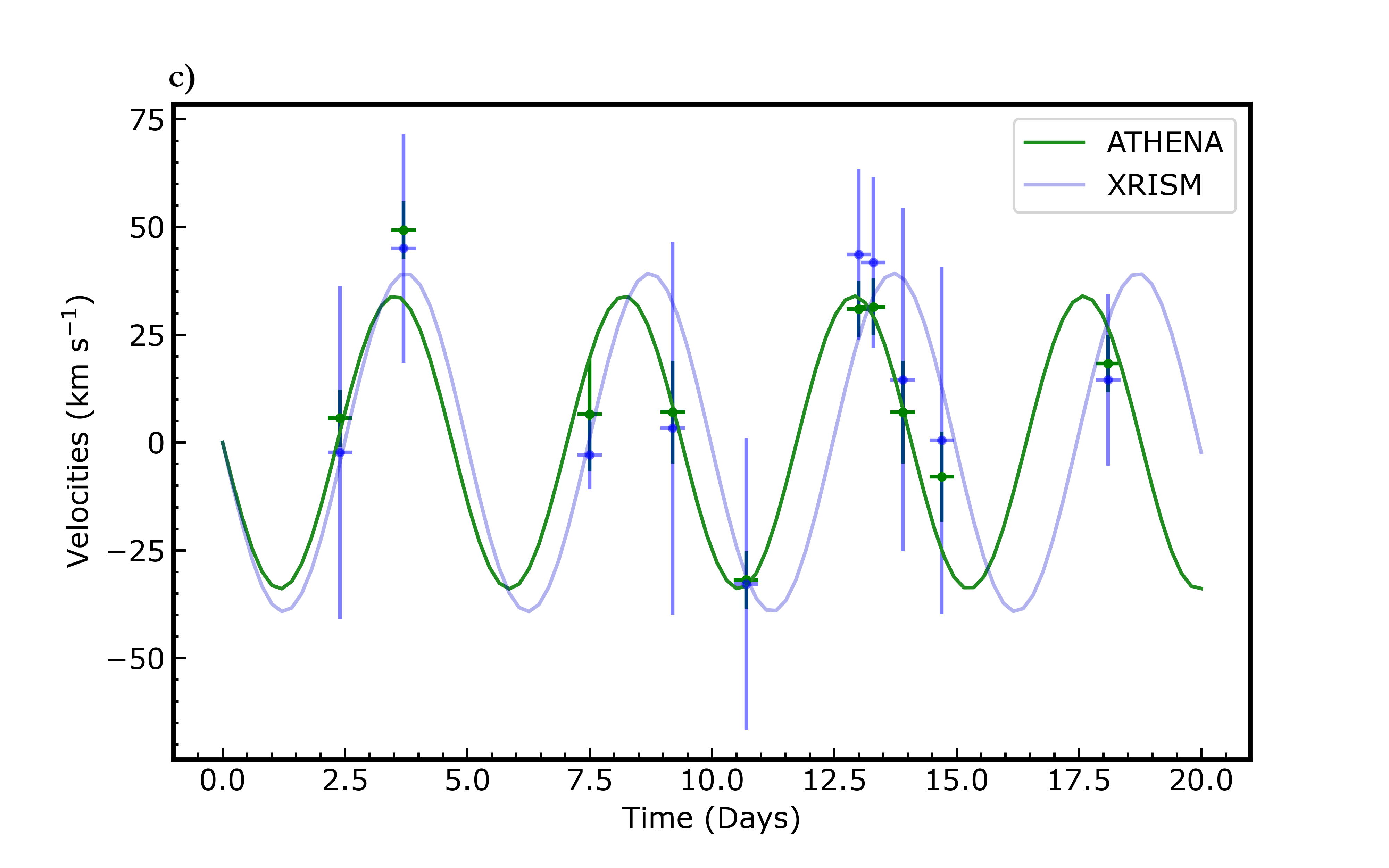}
\end{minipage}
\caption{Simulated radial velocity curves for a mass ratio $q$ = 4, assuming a flux of $10^{-10}$\,erg\,cm$^{-2}$\,s$^{-1}$, and for $K_2$ values of ({\em a}) 100\,km\,s$^{-1}$ and ({\em b}) 30\,km\,s$^{-1}$.  {\em Athena} and {\em XRISM} are compared ({\em c}) for the smaller $K_2$ case, indicating potential future improvements. Individual exposure times are 10ks, and the horizontal error bars in $x$ are 20ks.}
\label{fig:simradvel}
\end{figure}

We considered two HXMBs, each with a $q$ of 4 and the same $P_{orb}$, but with different $K_2$ values. Having found the expected EW of the K$\alpha$ line as a function of phase, we reduced this by 20\% to account for the changes in flux at the peak line energy due to the complex profile of the emission line (the Compton Shoulder, resulting from the down-scattering of fluorescent photons). 10 observations were chosen randomly over 4 orbital cycles, the Keplerian velocity of the K$\alpha$ line was calculated, and some scatter was introduced (mimicking small velocity changes associated with winds etc.). Simulations were then carried out to determine how well {\em XRISM} would be able to fit these spectra given the expected equivalent widths, and the radial velocity measurements and errors we might be able to produce. The results of these are shown in Fig.\,\ref{fig:simradvel}.

Where the flux of the X-ray source is $10^{-10}$\,erg\,cm$^{-2}$\,s$^{-1}$, errors in velocity range from 10 to 30\,km\,s$^{-1}$  (changing as a function of phase in response to the modulation of EW). For systems with a $K_2$ of order 100\,km\,s$^{-1}$ , these errors are small enough to produce good radial velocity curves, with fits in good agreement with the 'correct' system parameters (Fig.\,\ref{fig:simradvel}a). Where Keplerian velocities are smaller, $\sim$30\,km\,s$^{-1}$  (i.e. as would be more typical for systems with larger companion star masses), these errors cast doubt on the resultant radial velocity curves (Fig\,\ref{fig:simradvel}b). Fig.\,\ref{fig:simradvel}c shows a comparison between those produced with {\em XRISM} and {\em Athena} response files - indicating the significant improvement that {\em Athena} will provide, particularly for fainter fluxes and/or smaller velocities.

\section{Discussion}
These calculations indicate that, if narrow, fluorescent X-ray emission lines from the companion star can be isolated, they provide a viable method for measuring binary mass functions. This then makes it possible to investigate many systems that are too obscured and/or faint to be studied in the optical. 

The most viable candidates are those systems (mostly HMXBs) where the companion star presents a significant reflecting surface to incident X-rays, i.e., the line EW increases with mass ratio $q$, and where their inclinations are > $60^{\circ}$, (particularly if the mass ratio is smaller).  This needs to be balanced against the fact that more massive systems will have lower $K_2$ values, and there could be other, more complex, emission components such as stellar winds, as we discuss below. Hence, the ideal systems could be objects with more intermediate mass ratios $q$ and mid-spectral types. Exposure times of 100\,ks yield sufficient energy precision with {\em XRISM}, although shorter observations can provide useful constraints, especially with {\em Athena}.  Additionally, full orbital phase coverage is essential.

After the launch of {\em XRISM} and {\em Athena}, we recommend initial validation observations of systems with already well-established optical ephemerides, e.g. GRO\,J1655--40 ($q$\,=\,0.4, $P$\,=\,2.6\,days, $K_2$\,=\,228\,km\,s$^{-1}$), Cyg\,X\,-\,1 ($q$\,=\,1--2, $P$\,=\,5.6\,days, $K_2$\,=\,76\,km\,s$^{-1}$), or V404\,Cyg ($q$\,=\,0.1, $P$\,=\,6.5\,days, $K_2$\,=\,210\,km\,s$^{-1}$).\footnote{System parameters from \citet{Casares2015}.} These will provide a wealth of information on the viability of isolating the narrow component of the Fe line and its location, since the binary solution is known from the optical. In the cases of the two X-ray transient LMXBs (GRO\,J1655--40, V404\,Cyg), observing in {\em quiescence} is also potentially valuable, as it would minimise potentially confusing emission from the (still present) accretion disc. Additionally, during quiescence, discs may have lower flaring angles, thereby increasing the irradiated area of the companion \citep{Davanzo2006}. This would, however, be challenging, as the X-ray fluxes would be lower. If velocity errors from observations of K$\alpha$ are too large, other spectral features at lower energies (O, Ne, Fe-L at 0.5--1\,keV) associated with intrinsic coronal emission from the donor, might also be detectable.

Two further XRB candidates of particular interest for an X-ray spectroscopic study include Cyg\,X-3, and Her\,X-1. Cyg\,X-3 is an optically obscured, short period (4.8\,hrs), HMXB, comprising a (candidate) black hole and a Wolf-Rayet donor.  Even though it is X-ray bright, Cyg\,X-3's ephemeris is well established, and so phase-binning could be used to search for weak, sharp features within the spectra.  Her\,X-1 is eclipsing (1.7 day period), and it exhibits a well-known 35 day super-orbital periodicity \citep{LeahyIgna}, which is attributed to the tilting and precessing of its accretion disc, which causes variable X-ray obscuration.  Her\,X-1 is also an X-ray pulsar, and so it falls into the rare category of "intermediate mass X-ray binaries".

Our findings of an expected equivalent width of 5--40\,eV are in good agreement with the first calculations of this effect \citep{Basko1978}; e.g., for Her\,X-1 with a mass ratio $q$\,=\,1.8, we calculate an EW of $\approx 16$\,eV, very close to \citeauthor{Basko1978}'s estimate of 14\,eV.  Interestingly, \citeauthor{Torrejon2010}'s 2010 survey of K$\alpha$ emission finds EWs (particularly for HMXBs) are observed to be much higher, often a few hundred eV. This is likely due to the origins of the K$\alpha$ emission in HMXBs being the stellar wind - if not confined to one side of the binary system, the winds present a large solid angle through which X-ray reprocessing and therefore K$\alpha$ emission may occur. 

\subsection{Caveats}

Having presented our simulations and expectations, we now discuss some of the limitations, caveats and assumptions that need to be kept in mind. 

\subsubsection{Resolving and Locating the Narrow Line Component}
This work depends on our ability to isolate the narrow component of the iron fluorescent emission. We are cautiously optimistic that this will be feasible with microcalorimeters, as a narrow-line component has been isolated in previous studies by {\em Chandra} HETG at the best spectral resolution currently available in the Fe K band ($\Delta E$\,=\,40\,eV). For comparison, some of the previous best studies with CCDs, using {\em XMM-Newton}, have found narrow features in the X-ray spectra of HMXBs, but with uncertainties in centroid energy of order 10\,eV -- corresponding to velocity errors $>$200\,km\,s$^{-1}$ (and is consistent with estimates from XSPEC simulations - see Fig.\,\ref{fig:contour}b) \citep{Garcia2015}. We expect the gains with {\em XRISM} (and later {\em Athena}) to be such that we will not only be able to identify the narrow line in a number of systems, but also to obtain velocity information concerning these binary systems.

We must also point out that, while we have taken the companion as the most likely site for a narrow K$\alpha$ component, other locations are known to be possibilities e.g. stellar winds, inner accretion disc structure, the impact bulge due to the accretion stream.  If present, such components would make it harder to determine system masses. But, using both velocity and flux variability information as a function of phase would allow them to be physically located within the binary system, thereby allowing the detailed accretion geometry in and around the disc to be constrained in ways not hitherto possible.  (See also Section 5.2)

Whilst our simulations treat the fluorescent emission as a simple Gaussian, for demonstration purposes, the actual line profile of the narrow component will likely be more complex. The high energy fluorescence photons may have some probability of being Compton down-scattered to lower energies, resulting in a "red shoulder" in the K$\alpha$ line profile \citep{Watanabe2003}.
This has been observed in some HMXBs with supergiant companions  \citep{Torrejon2010}, and we have tried to account for this by reducing the EW of the Gaussian peak by 10\% when performing the more detailed radial velocity simulations in Section 4.1. Models of the line's complex structure suggest that the central line energy should still remain constrained, and the peak flux will be 60--90\% of the 'expected flux'. Attention to complexities in the line profile should be noted, both as a caveat to the accuracy of our predicted emission strengths and radial velocity measurements, and as an opportunity to further understand the reprocessing material around the XRB.

\subsubsection{Stellar Metallicity}
The K$\alpha$ line's EW is dependent on the stellar metallicity, and spectra are usually fitted with solar-metallicity models. Given that the disc is accreted from a companion likely to have close to solar-abundances, it is not surprising that previous studies into XRB metallicities find iron abundances similar to, or slightly exceeding solar values (i.e. [Fe/H]$\approx$0.24 in NS binaries, and [Fe/H]$\approx$0.09 in BH binaries (\citealt{Gonzales2011} and references therein). Higher metallicity is not unexpected, as the companions are likely to be highly evolved. Additionally, there is evidence  that the formation of the compact object via a supernova explosion can lead to significant amounts of processed material being captured by the companion star, thereby leading to higher metallicities \citep{Shahbaz2022}. It is, therefore, plausible that the companion star's atmosphere will have sufficient iron content to produce X-ray fluorescence. Nevertheless, this needs confirmation, and such studies will provide information about the donor's evolution.

\subsubsection{Companion Star Mass}
This work has focussed on using the K$\alpha$ line velocities to constrain $f(M_1)$.  However, while HMXB donors present a much larger solid angle to the X-ray source (assumed to be a NS, as they dominate the HMXB population), the donors will typically have lower velocities ($\sim$ a few 10\,km\,s$^{-1}$ ), which may be comparable to their accompanying errors, making meaningful radial velocities difficult to extract. Fig\,\ref{fig:simradvel}b demonstrates this effect, which is of course flux-dependent. We hope that the larger donor solid angle and resulting stronger K$\alpha$ emission in such systems means that the velocity errors will be better constrained. Alternatively, on a longer timescale, {\em Athena} will have sufficient resolution and greater sensitivity, with errors decreasing from order 15\,km\,s$^{-1}$  to  $\sim$5\,km\,s$^{-1}$  at $F$\,$<$\,10$^{-9}$\,erg\,cm$^{-2}$\,s$^{-1}$.  

Exposure times for our simulations ranged from 5\,ks to 50\,ks. For short period XRBs, phase-binning will be required, and the periods are expected to be independently known. 

\subsubsection{Complexities related to wind dynamics}
One of the biggest uncertainties associated with this methodology is the fact that the nature and impact of the stellar winds are still somewhat uncertain. There have been many studies of wind structure in XRBs, with strong effects (particularly large X-ray variability) indicating the presence of clumping \citep{Torrejon2015}. This leads to deviations from the expected radial velocity curves, with the effects being most prominent near phase 0, i.e. inferior conjunction \citep{Grinberg2015, Hirsch2019, Lae2022}. Previous studies of lower-energy lines (such as silicon and sulphur) have been used to study the stellar winds in Cyg\,X-1 \citep{miskovicova2016}, but earlier missions have lacked the high-energy resolution to extensively study the K$\alpha$ line in the same way. 

With sufficiently long observing times, the Doppler scatter due to clumpy winds will average out, and details of the complex wind structures will be of interest regardless. \citet{Torrejon2010} did not in fact detect any substantial changes in the {\em peak line-energy} of their detected narrow line components that might have been expected due to winds, which provides encouraging support for our proposed study. Instead, the maximum width ($\sigma$) of the unresolved narrow line components (5\,eV) placed a constraint on a plausible wind velocity of $\nu$\,=\,800\,km\,s$^{-1}$. {\em Chandra}'s resolution meant that line centroid energies in that survey were constrained with errors of 6\,eV, corresponding to 250\,km\,s$^{-1}$. We can therefore speculate that either the winds do not inflict random scatter in the peak energy of the K$\alpha$ emission, or that scatter of this type is of order $<$\,250\,km\,s$^{-1}$ . This may still be comparable with orbital velocities, and so it will only be known for certain how much of an impact this will have once observations are carried out. 

LMXBs generally have much weaker donor wind components, and so velocity measurements should be less susceptible to variations that are not associated with donor orbital motions. Indeed, the success of \citet{Ponti2018} in constraining the mass function through the radial velocities derived from absorption features associated with the compact object indicate that whilst more complex physics may be at play, this does not necessarily preclude X-ray mass function measurements. We highlight again our suggestion to test this technique on some  well-characterized system (such as GRO\,J1655--40) in order to determine whether the stellar wind presents a significant and/or systematic hindrance.

\subsubsection{Ionisation Effects}
If there is any change in the ionisation state of the reflecting surface, the energy of the fluorescent emission will change. This would cause a shift in peak energy unrelated to orbital dynamics. Previous studies have used spectra of the fluorescent emission to determine the ionisation states of the reprocessing material (e.g. \citealt{Garcia2015} found that the peak energy of the Fe K$\alpha$ fluorescent component was centred on 6.42\,keV, and suggested that the reprocessing material must be in an ionisation state below Fe {\sc xviii}). Variations in the ionisation level and therefore peak energy could complicate radial velocity measurements. 

If X-ray sources become transiently bright or dim, this could affect the ionisation of the stellar wind or companion star surface, resulting in variations in K$\alpha$ line energy. Additionally, the impact of the accretion stream on the edge of the disc results in a ``bulge'' structure, and it has been suggested that this can cause ionisation instabilities, creating patchy clouds of colder gas among the hotter medium \citep{Shidatsu2013}.
However, if these effects should be systematic and persist across the orbital cycle, this should not affect radial velocity curve measurements, since our main observable of interest is the Doppler variation with orbital phase. 

While this remains to be tested, observations in {\em quiescence} may be more favourable to avoid the complications associated with ionisation effects, as at the expense of a lower source flux, systems will likely be subject to fewer changes in ionisation and continuum flux during quiescence. Simulations demonstrate that observations of K$\alpha$ emissions even at low fluxes (10$^{-11}$\,-\,10$^{-12}$\,erg\,cm$^{-2}$\,s$^{-1}$) could be used to constrain velocities to $\pm$30\,km\,s$^{-1}$. This will be challenging, but would, obviously, be more effective with {\em Athena}, given its larger collecting area.

\subsubsection{Instrumental Systematics}

We have focused on a proof-of-concept here, and largley ignored any instrument-specific biases, although we note that  microcalorimeters will be optimised for relatively faint sources. High count-rate observations may suffer from other systematic issues such as cross-talk \citep{Colas1999}, which would reduce the effective spectral resolution of some fraction of detected events at bright fluxes, thus adversely impacting radial velocity variation searches.

This is another reason to attempt to exploit fainter {\em quiescent} states for this work where possible. Ultimately, we also recommend that instruments on future facilities are developed to robustly handle high count-rates. \citet{Kammoun2022} propose that, when observing very bright sources, {\em Athena}'s X-IFU could make use of defocussing, where the telescope's point spread function (PSF) is spread over multiple pixels. In this way, spectra of bright sources could be successfully reconstructed, thereby mitigating some of the instrumental limitations.

We also note that, at the time of writing, descoping of {\em Athena} remains a possibility, which could obviously have a detrimental impact on mass function measurements. Our proposed science case adds support to retaining the full effective area of the microcalorimeter.

\subsubsection{Period Determination}
Finally, even in the best possible circumstances, exploiting these spectral data to extract radial velocity curves is a non-trivial task. 
Firstly, multiple observations are obviously required, and in most XRBs the orbital period is likely to be already known (through other data, usually optical, IR or radio), making broad phase coverage possible. One advantage with X-ray observations is the possibility of coadding multiple datasets into phase bins, allowing maximum sensitivity for searching for narrow spectral features.  

Furthermore, the number and cadence of observations will also impact the ease with which a radial velocity curve can be extracted. When searching for periodicities in {\em photometric} light curves, for instance, the influence of red noise cannot be ignored \citep{Vaughan2009}. Here, we are instead fitting {\em spectroscopic} data. Whilst the above simulations have relied upon being able to fit sinusoidal curves to the data, depending on the number and cadence of observations, the magnitude of statistical scatter in the data, and potential radial velocity scatter due to multiple intrinsic line components, a more detailed Fourier period search will likely be required \citep{Scargle1982, Horne1986}. 

These are important ultimate considerations. But our goal here has been to investigate the viability of detecting X-ray radial velocity changes in the first place; fitting the ensemble measurements will be source-specific, which is beyond the scope of the present work.

\subsection{Disc Origin and Inversion of the Mass Function}
Whilst this paper focuses on using the X-ray fluorescence from the companion star, there exists the potential for this method to be applied to reflection off disc emission instead (with some modifications). This would be relevant if the narrow line originates from the outer disc, or from a disc wind anchored to the primary. 
In such cases, the mass function is effectively inverted, and it may be possible to measure the Doppler shift in the emission from the accretion disc in order to constrain the mass of the companion star instead. If the mass function were inverted, we get;
\begin{equation}
    f(M_2) = \frac{M_2^3\mathrm{sin}^3i}{M_{Tot}^2} = \frac{P_{orb}K_1^3}{2\pi G}
\end{equation}

The masses of companion stars are also often uncertain, as binary interactions can complicate stellar evolution. Such an inversion could then provide an interesting and independent validation of companion star parameters.

Conversely, if we have an independent estimate of M$_2$ (for example, from its spectral type), we can use the radial velocity curve from the outer disc of the compact object to constrain the mass of the compact object. In this case, we solve for M$_1$ as below:
\begin{equation}
    0 = f(M_2)\times(M_1+M_2)^2-M_2^3\mathrm{sin}^3i
\end{equation}
\begin{equation}
    M_1 \leq \frac{\sqrt{M_2^3\mathrm{sin}^3i}}{f(M_2)} - M_2
\end{equation}

Both \citet{Ponti2018} and \citet{Zhang2012} use the Doppler motion of absorption features associated with the accretion disc to infer system parameters (i.e. companion star mass and black hole mass \& system inclination respectively), making use of prior observations and well-defined system parameters (i.e. \citet{Zhang2012} used the known phase of the companion to fit velocity curves for the accretion disc). That this method has been validated with {\em Chandra} data demonstrates the vast potential in X-ray spectroscopy as a way to probe system characteristics.

\section{Conclusions}
Having demonstrated the theoretical validity of using X-ray spectroscopy to produce radial velocity curves for binary systems, and discussed several of the main caveats to keep in mind, we propose that measurements of mass function be seriously attempted with upcoming X-ray missions. Our main findings can be summarised as follows.

\begin{enumerate}[(i)]
    \item The companion star in X-ray binary systems can present a large solid angle, particularly in systems where the mass ratio, $q$, exceeds 1. This leads to iron-line fluorescence with an expected equivalent width of the K$\alpha$ emission $\sim$\,2--40\,eV, for systems with $q$\,$>$\,0.1. 
    \item New microcalorimeters will be able to detect the K$\alpha$ iron line, and constrain velocities within $\sim$\,5--30\,km\,s$^{-1}$ . Given sufficient observations spanning their full orbital cycles, this should be sufficient to produce a radial velocity curve of the companion star in XRBs, and therefore measure the mass function of the compact object.
    \item This method is subject to a number of caveats and conditions, for example; the complexities within the stellar wind; low companion velocity; and the usual systematic constraints associated with radial velocity measurements. The narrow line component from the donor also remains to be unambiguously identified.
    \item Both LMXBs and HMXBs are viable targets, however there are limitations associated with each type of system. LMXBs are expected to have lower equivalent widths and therefore larger errors in velocity, however companions in HMXBs are more likely to have smaller radial velocities, which may make errors more significant.
\item High resolution observations of K$\alpha$ emission in XRBs may also develop understanding of the behaviour of the stellar winds in HMXBs.
\end{enumerate}

\vspace*{0.5cm}
\noindent
Despite the above caveats, the potential gain in parameter space is huge, enabling new system constraints on previously inaccessible sources (that are either too faint, obscured, or otherwise difficult in the optical). The new era of high resolution X-ray astronomy will be a step-change in terms of improved precision constraints on binary system parameters. These advances should be exploited to address fundamental questions regarding the mass distribution of compact objects and related aspects of stellar evolution.

\section*{Acknowledgements}

This work is funded by STFC grant reference ST/V001000/1. We would like to thank A. Veledina, J. Casares, A. Zdziarski, and M. Diaz Trigo for helpful comments and corrections. We are grateful to V. Grinberg and members of the {\em NuSTAR} Binaries science team for very useful discussions. In particular, J.\,M. Miller helped with additional useful references and food-for-thought. We also thank our referee for their feedback and insight. 

\section*{Data Availability}

No new data presented in this work. Simulation files are publicly available on respective observatory archives.



\bibliographystyle{mnras}
\bibliography{refereneces} 


\appendix


\bsp	
\label{lastpage}
\end{document}